\documentclass[twocolumn,prl,showpacs,floatfix]{revtex4}

\usepackage{graphicx}
\usepackage{epsfig}
\usepackage{rotating}
\usepackage{amsmath}
\usepackage{amsfonts}
\usepackage{amssymb}
\usepackage{enumerate}
\usepackage{longtable}
\usepackage{dcolumn}
\usepackage{bm}

\begin{document}

\title{Ground state and spectral properties across a
Charge Density Wave transition in a triangular-lattice spinless Fermion model}

\author{Trithep Devakul$^1$ and Rajiv R. P. Singh$^2$}
\affiliation{
    $^1$Department of Physics, Northeastern University, MA 02115, USA \\
$^2$Department of Physics, University of California Davis, CA 95616, USA}

\date{\rm\today}

\begin{abstract}
We study ground state properties  and particle excitation spectra across a commensurate charge density wave
transition in a system of strongly interacting fermions, using series expansion methods and mean-field theory.
We consider a $1/3$-filled system of spinless fermions on a triangular-lattice,
with hopping parameter $t$, nearest-neighbor repulsion $V$, and a sublattice dependent chemical potential $\mu_s$.
The phase transition is found to be first order for $\mu_s=0$, but becomes continuous with increasing $\mu_s$.
The particle and hole excitation spectra exhibit dramatic changes in the vicinity of the phase transitions and
in the charge-density wave ordered phase.
We discuss the relevance of this study to
the Pinball Fermi liquid phase postulated theoretically in earlier studies
as well as to various strongly correlated triangular-lattice materials.
\end{abstract}

\maketitle

\section{Introduction}

Quantum oscillations and evolution in shapes of Fermi surfaces with various parameters have been
recognized as key issues in studies of high temperature superconducting cuprates and other
strongly correlated materials \cite{sudip}.
In general, these systems show many coexisting and competing order parameters and collective fluctuations.
A number of theoretical studies \cite{sudip2,palee,kivelson,kivelson2,vojta,berg} have shown that several such orders leading to broken symmetry,
such as spin-density waves, charge-density waves, $D$-density waves or pair-density waves cause a partial
or complete gapping of the Fermi surface with substantial, in some cases even topological, changes in the Fermi surface
geometry in the transition region. In the cuprate family of materials, the role of such changes in Fermi surfaces
in providing a quantitative understanding of the quantum oscillation measurements remains a hot topic \cite{sudip}.
The interplay of these orders with high temperature superconductivity, non-fermi liquid behavior and pseudogap
also remain subjects of intense current research \cite{melko}.

Controlled microscopic calculation of properties of strongly correlated electron systems, such as Hubbard and t-J models,
remains a challenging task. While Quantum Monte Carlo methods remain the most powerful for some models \cite{scalapino},
in many cases they suffer from `minus-sign' problems, which
severely restricts simulations at low temperatures. Recently developed variational methods that build in low ground-state entanglement,
such as density matrix renormalization group \cite{dmrg} and tensor-network methods \cite{tensor} are promising, but much progress remains to be achieved.
High temperature expansions \cite{hte} and Numerical Linked Cluster methods \cite{nlc} can obtain temperature dependent
properties in the thermodynamic limit, but
have found it difficult to address ground state and low temperature properties of these systems. Much quantitative
information on Hubbard and t-J models still comes from exact diagonalization of rather small systems \cite{ed}.

In this paper, we focus attention on charge density wave (CDW) order in a triangular-lattice system relevant to a number of organic molecular
crystals as well as to the sodium cobaltate family of materials \cite{motrunich,ogata,powell,mckenzie,cava-naco}. 
It has been suggested that over a wide range of electron density,
these systems may be described as a Pinball liquid, where there is a background commensurate CDW 
order\cite{footnote1} on the triangular-lattice corresponding to a $1/3$ or $2/3$ filling, while the rest of the electrons or holes form a 
metallic fermi-liquid that move in the background of such an order \cite{nishimoto,hotta,hotta2}.
Furthermore, it has been argued that for very strong on site Hubbard repulsion $U$, where double occupancy is strongly disfavored,
it may suffice to model the charge fluctuations in the system in terms of spinless fermions \cite{hotta,hotta2}. 
In this paper, we will study this specific model
at a commensurate $1/3$ filling. While the extension to spinful fermions is conceptually straightforward, the much larger Hilbert 
space of the spinful problem could severely limit the efficacy of series expansions.
We also find that fairly high orders in the series are necessary for convergence even further away from the critical point.

Our results are based on two types of exact zero-temperature series expansions \cite{book,series-reviews}, carried out to high orders, as well as
a mean-field treatment of the electron-electron repulsion. Unlike the square-lattice case, where nesting causes order to develop at
infinitesimal $V/t$, on the triangular-lattice, at small $V/t$, the ground state of the system remains a Fermi liquid.
Only beyond a critical $V/t$ one obtains a commensurate CDW order. In the latter phase, we expand ground state and low 
lying spectral properties in powers of $t/V$. For a second expansion we introduce a sublattice-dependent chemical potential $\mu_s$, which
favors occupancy of one sublattice over the other two. In a real material, such a term can arise from having two types of atoms, 
or from two different local environments in a unit cell.
This allows us to carry out a different series expansion in inverse
powers of the chemical potential. The chemical potential term can also arise from a self-consistent mean-field treatment of
the charge-density wave order and this latter approach allows us to obtain a more complete picture of the evolution of the fermion excitation
properties in the transition region.

The picture that emerges for the electronic states across the transition is a very rich one. If we consider the non-interacting system with $V=0$ and
examine it as a function of $\mu_s/t$, we find that 
the gap first opens at `hot spots' which correspond to the points in k-space where the non-interacting Fermi surface crosses the reduced
Brillouin zone, analogous to many recent modeling of the cuprates \cite{berg}. This creates a topological change in the Fermi surface breaking it up
into many alternating electron and hole like pockets. Eventually, when a full gap opens, the system becomes an indirect
gap charge density wave semiconductor. However, the $\mu_s=0$ system shows a very different behavior.
Coming from the CDW side in self-consistent mean-field theory, we find that as the indirect band-gap is closing, the order parameter jumps to zero. 
Thus, one obtains a discontinuous change
from a non-interacting fermi surface to a gapped phase. One should note that the CDW order in
our system has the symmetry of a 3-state Potts model and in $3$ and higher dimensions the latter model is known to have
a weakly first order transition \cite{potts}.

At small $t/V$ and $\mu_s = 0$,
the low energy particle excitations are first centered around the $\Gamma$ point.
However, as $t/V$ is increased a more prominent minima develops near the edges of the reduced Brillouin zone. 
As the transition away from the charge density wave phase
is approached, the particle hole excitation becomes soft
all across the reduced Brillouin zone boundary. We find good agreement from the results of the two different series expansions.
We also find that adjusting a parameter in the self-consistent mean-field theory can make it quantitatively accurate
for the ground state energy and the charge density wave order parameter.

The spectra at small $t/V$ are relevant to the theoretically postulated Pinball liquid phase away from $1/3$ filling. 
At half filling for a system of spinless fermions, the Pinball liquid phase was found to be stable only for
small $t/V \le (t/V)_c$, where $(t/V)_c$ was estimated to lie between $1/12$ and $1/5$.\cite{nishimoto,hotta,hotta2}
At larger $t/V$ values there was a phase transition to stripe-order and other phases. It is clear that
the Pinball liquid phase would be stable to larger values of $t/V$ as the particle density is reduced towards $1/3$, eventually
at $1/3$ filling the transition happens at the mean-field value obtained in this study $t/V \approx 0.49$. In a rigid-band picture, the additional particles beyond $1/3$ filling
would occupy the particle excitation bands we have calculated. Our results should become exact as the doping beyond $1/3$ filling
goes to zero.

Our study is possibly relevant to organic molecular crystals as well as to the sodium cobaltate family of materials.
Charge order with $1/3$ or $2/3$ occupation have frequently been suggested but not clearly 
established \cite{motrunich,ogata,nishimoto,hotta,hotta2} in these materials.
One advantage of the former system is that they are very pressure sensitive and the degree of strong correlation can
be changed continuously by pressure thus driving the system across various quantum phase transitions. However, it is difficult to
change the carrier concentration in these materials, which are generally quarter or half filled. 
In contrast, in the sodium cobaltate family, doping can
be achieved by various substitutions as well as by water intercalation. There have also been meaurements
of angle resolved photoemission spectra for the latter materials \cite{naco-arpes,naco-arpes2,naco-arpes3,naco-arpes4,naco-arpes5}. 
However, although theoretically postulated, such 
a charge density wave state has not been convincingly seen in experiments. When such a transition is present, 
our work can form the basis for understanding the fermion spectra.
Our results are most directly applicable to a spin polarized system or an artificially engineered cold atomic gas system
with only one fermionic spin species. They also provide an example where density wave induced changes in fermi surface 
properties and limitations of mean-field approaches can be systematically explored.


\section{Method}

\begin{figure}
\begin{center}

\includegraphics[width=0.35\textwidth]{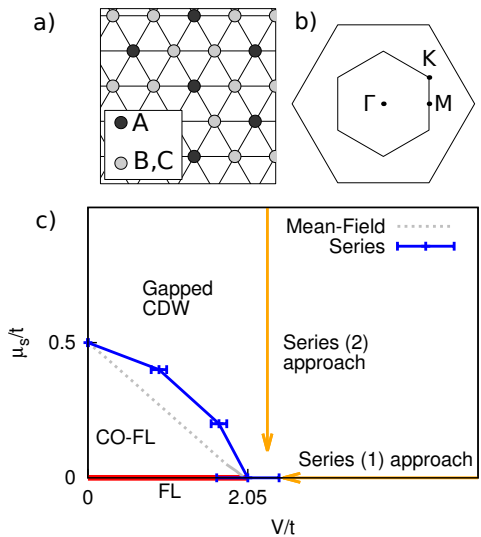}
\caption{\label{phasediagram}
(a) A diagram of the three A-B-C sublattice structure of the CDW phase.  In the unperturbed ground state, the 
sites of A sublattice (dark circles) are occupied, and the B and C sublattices are unoccupied.  
(b) A diagram of the reduced Brillouin zone inside the full triangular-lattice Brillouin zone.  The points $\Gamma$, M, and K in the reduced zone are labeled.
(c) Phase diagram, with the magnitude of the chemical potential $\mu_s$ on the vertical axis, and the repulsion $V$ on the horizontal axis.  
Arrows show the limit from which the expansions (1) and (2) begin, and the direction in which the series extrapolations are done.  
At low $V/t$ and $\mu_s=0$, there exists a Fermi liquid (FL) phase.  
At the introduction of a nonzero $\mu_s$, one has a charge-ordered Fermi liquid (CO-FL) phase at low $V/t$.  
At higher values of $\mu_s/t$ or $V/t$, and one has an insulating gapped CDW phase.
The phase boundary is calculated using (scaled) self-consistent mean-field theory and series expansions.
In the mean-field results, 
the dotted line indicates a continuous second order transition, and the solid line indicates a first order transition (discussed in text).
At $\mu_s=0$, series does not allow the determination of a first order transition, which is represented by the large error bars.
}
\end{center}
\end{figure}

The interacting spinless fermion Hamiltonian in a sublattice-dependent potential is given by
\begin{eqnarray}
    \mathcal{H} = V\sum_{\langle i, j \rangle} n_i n_j -t\sum_{\langle i, j \rangle}\left( c_i^\dagger c_j + c_j^\dagger c_i \right) + \sum_i \mu_{s}^i n_i
    \label{ham}
\end{eqnarray}
where $c_i^\dagger$,$c_i$ are the usual creation and annihilation operators, $n_i = c_i^\dagger c_i$ is the density operator, and the sum over $\langle i,j \rangle$ is over neighboring sites.
$t$ is a parameter which represents the kinetic energy of the fermions, and $V$ represents the nearest neighbor repulsion.
Fig \ref{phasediagram}a shows the three sublattice structure of the triangular-lattice.  
The $\mu_{s}^i$ is the sublattice-dependent chemical potential, which is taken to be $-\mu_s$ on the A sublattice, and $+\mu_s$ on the other two.
We consider only $\mu_s\geq0$.
As the limit $\mu_s\rightarrow 0$ applies to the system with no inherent broken symmetry, it will be a key focus of our study.
To study the system at the commensurate $1/3$ filling by series expansions, we take two different approaches.

For the first approach, we consider equation (\ref{ham}) with $\mu_s=0$.
We use as an unperturbed Hamiltonian the nearest neighbor repulsion, which has a simple three sublattice CDW ordered ground state, where
the A sublattice is fully occupied and the B and C sublattices are unoccupied.
We then treat the hopping term perturbatively, and expand in powers of $t/V$.
This expansion provides a natural way to study the system at $\mu_s = 0$.
This expansion is also done at non-zero $\mu_s/V$ or $\mu_s/t$, by simply adding the $\mu_s$ term to the unperturbed or perturbing Hamiltonian, respectively.

For the second approach, we treat $\mu_s$ as the unperturbed Hamiltonian, with the same CDW ground state.  Then, the Hamiltonian can be written
\begin{eqnarray}
    \mathcal{H} = \sum_i \mu_s^i n_i + \lambda \sum_{\langle i, j \rangle} \left[-t \left( c_i^\dagger c_j + c_j^\dagger c_i  \right)+ V n_i n_j  \right]
    \label{mu_exp}
\end{eqnarray}
and expanded in powers of $\lambda$ at any ratio of $t/V$.  
This can also be used to examine the system in the limit of $\mu_s \rightarrow 0$, or equivalently the asymptotic behavior as $\lambda \rightarrow \infty$.
This is done by a transformation of variables to $x = \lambda/\left(\lambda+1\right)$, which shifts the limit to $x = 1$.
The series can then be analyzed by Pade or differential approximants \cite{book,series-reviews}, and evaluated at $x=1$.

In Fig \ref{phasediagram}c, we show a phase diagram for this system.
We also show the approaches of the two series expansions based on equations (1) and (2), in working with the $\mu_s = 0$ system.
Along the $\mu_s/t$ axis, where $V = 0$, we have a system of noninteracting fermions, which can be solved through Fourier transformation and diagonalization of a $3\times 3$ matrix at each $k$.

We also carry out a self-consistent mean-field calculation for the system.  The mean-field acts effectively as a chemical potential on each sublattice on top of $\mu_s$, given by
\begin{eqnarray}
    \mu_A &=& 6V\langle n_B \rangle \\
    \mu_{B,C} &=& 3V \langle n_A \rangle + 3V \langle n_B \rangle,
\end{eqnarray}
where we have taken $\langle n_B \rangle = \langle n_C \rangle$.
This becomes a non-interacting problem, and self consistency of the sublattice occupancy density can then be enforced.
Quantities of interest can then be calculated in a straightforward manner.

With series expansions, in addition to ground state properties, we calculate the single-particle spectrum and 
the particle-hole excitation-gap.
We consider a single particle or hole excitation, corresponding to adding or removing a particle.
For each case, we obtain the effective Hamiltonian for the single-particle excitations, from which the excitation spectrum 
is readily obtained \cite{book}. 
As the particle (hole) energies are shifted up (down) by a uniform chemical potential, the particle-hole energy gap is the sum of the 
particle and hole excitation energies.
For both types of expansions, all ground state properties are calculated to order 12, except the spectra
and gap are calculated to order 11.

\section{Results}

\begin{figure}
\begin{center}
\includegraphics[width=0.23\textwidth]{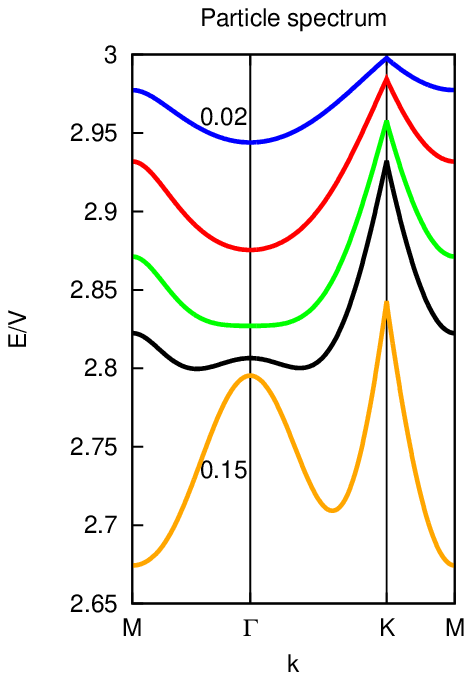}
\includegraphics[width=0.23\textwidth]{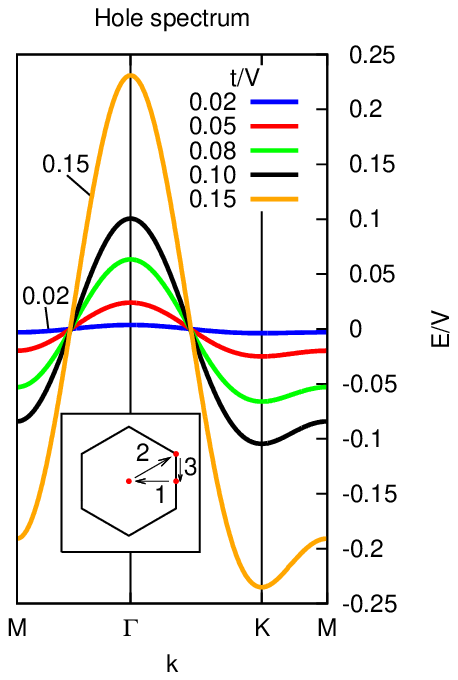}
\caption{\label{spectrum}
Particle and hole excitation spectrum at low $t/V$ ratios, with $\mu_s = 0$.  Only the lower of the particle excitation eigenvalues within the reduced Brillouin zone is shown, relevant to the Pinball liquid phase.
Here, our results should be highly accurate.
The particle spectrum begins free-particle like at low $t/V$, 
then quickly develops a minimum at the M point.  
The hole spectrum is always at a minimum at the K point.
As one increases $t/V$ further, the gap will begin to close between the M point of the particle spectrum and the K point of the hole spectrum.
}
\end{center}
\end{figure}

\begin{figure}
\begin{center}
\includegraphics[width=0.4\textwidth]{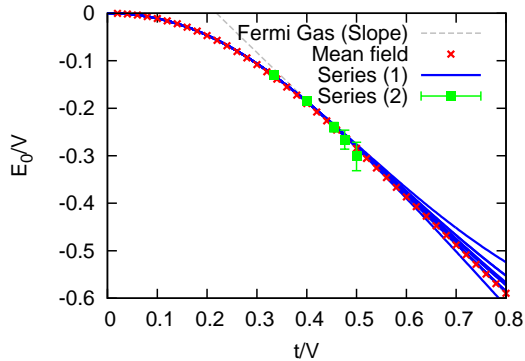}
\caption{\label{energy}
Ground state energy per site as a function of $t/V$, at $\mu_s = 0$.  
Results for series based on equation (1) and (2) are shown.
For the series (1), we show some well-behaved integrated differential approximants, whose spread determines the precision of our results..  
Our series (2) errorbars show the spread in the values obtained from Pade approximants in the $\mu_s\rightarrow 0$ limit.
The mean-field result has the interaction energy scaled by $V \rightarrow 2V/3$.  
After this somewhat arbitrary scaling, mean-field theory produces visibly accurate numerical results.
All subsequent mean-field results have been scaled in this manner.
Also shown is the slope of the noninteracting Fermi gas, to which mean-field theory has a first order transition to at $t/V\approx0.49$, shifted by the (scaled) mean-field interaction energy $2V/9$.
Here, both series show good convergence.
}
\end{center}
\end{figure}

\begin{figure}
\begin{center}
\includegraphics[width=0.4\textwidth]{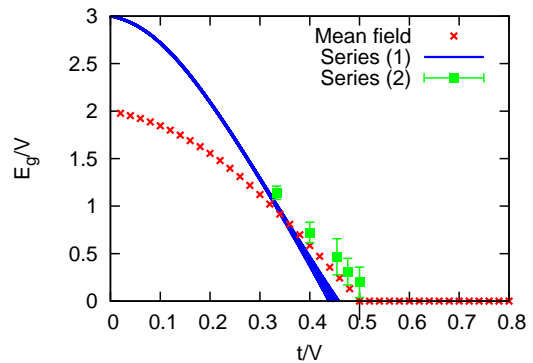}
\caption{\label{gap}
Particle-hole excitation gap defined as a sum of the minima of the particle excitation energy at M and
the minima of the hole excitation energy at K, with $\mu_s = 0$.
Well-behaved integrated differential approximants are shown for series (1).
The scaled self-consistent mean-field theory results are also shown.
The mean-field solution indicates that the gap closes discontinuously at $t/V\approx0.49$, very close to where it would close naturally in the absence of a first order transition (not distinguishable in plot).
Mean-field theory obtains incorrect values for the gap, even at low $t/V$, due to the scaling of $V$.  
}
\end{center}
\end{figure}

\begin{figure}
\begin{center}
\includegraphics[width=0.4\textwidth]{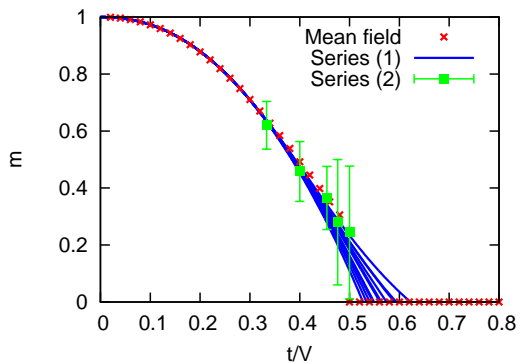}
\caption{\label{orderparam}
Order parameter as defined in equation (\ref{opdef}) as a function of $t/V$, at $\mu_s = 0$.  
As is the case with the energy, there is remarkable agreement between the series and (scaled) mean-field results, which suggests that mean-field theory is numerically accurate, but overestimates the interaction energy.
At $t/V\approx0.49$, the mean-field solution undergoes a first order transition to the Fermi liquid phase, which 
corresponds to the discontinuity in the order parameter.
Well-behaved integrated differential approximants are shown for series (1).
}
\end{center}
\end{figure}

We first focus on the $\mu_s = 0$ case, where there is no inherent symmetry breaking.  
In the low $t/V$ region, a Pinball liquid state has been studied \cite{nishimoto,hotta,hotta2}.
Here, fermions on the fully occupied sublattice remain insulating, which serve as a background through which any additional fermions beyond $1/3$ filling move freely around.  
Indeed, as can be seen in Fig \ref{spectrum}, the single particle excitation spectrum at low $t/V$ resembles that of free fermions.  

At $t/V = 0.08$, we begin to see an inversion of the spectrum.  By $t/V = 0.15$ a clear minimum has developed at the M point, which remains the minimum for higher $t/V$.  
Meanwhile, the hole spectrum displays a minimum at the K point.
At the critical point, where one has gapless excitations, one expects that the sum of the two energies at these points will become zero.

As one further increases $t/V$, there is a transition from the gapped CDW phase to a gapless Fermi liquid phase.
This transition has the same symmetry as in the three-state Potts model, where there is a first order transition from a three-fold degenerate
symmetry broken phase \cite{potts}. 
Thus, we also expect that this transition is also first order.
To identify a phase transition, we additionally compute the order parameter, defined as
\begin{eqnarray}
    m &\equiv& \langle n_A \rangle - \langle n_B \rangle 
    \label{opdef}
\end{eqnarray}
where we have again assumed $ \langle n_B \rangle = \langle n_C \rangle $.
This order parameter is thus at a maximum of 1 at $t = 0$, and becomes 0 in the Fermi liquid phase.

We first discuss the series results, which one expects to be more accurate than mean-field theory.  
When expansions from only one side is possible, series methods are unable to locate first order transitions.  
However, series can give good results up to the transition point, and can serve as a solid numerical ground to which mean-field theory can be compared.
Results from both types of expansions (1) and (2) show excellent agreement for the energy (Fig \ref{energy}).
In Fig \ref{orderparam}, series approximants for the order parameter are shown.  
However, as the order parameter does not go continuously to zero across the first order transition, it cannot be used to identify the 
transition point.  

To evaluate the gap series from expansion (1), we improve convergence by first a transformation of variables to $x = {\lambda}/{(1+\lambda)}$, where $\lambda$ is the expansion variable.  
Integrated differential approximants can then be used to evaluate the series to accurately determine the point at which the gap closes.
They suggest that the gap closes at $t/V=0.447 \pm 0.005$.
The results of series (2) expansions suggest that the gap closes at a slightly higher point.
While the gap closing does not necessarily indicate the transition point,
a simple argument that it should closely correspond to the first order transition point will be presented.

Next, we remark on the self-consistent mean-field results.
Alone, the results of such a mean-field approach is expected to be only qualitatively correct.
However, with our series results, we find that remarkably consistent numerical results can be obtained from the mean-field approach by simply scaling $V \rightarrow 2V/3$.
In Fig \ref{energy}, we show results for the ground state energy per site.
After scaling by this ad-hoc factor of $2/3$, we find that the mean-field results show remarkable numerical agreement with the series results.
The same is also true for the order parameter, shown in Fig \ref{orderparam}.
In other words, mean-field theory is found to overestimate the interaction energy by a factor of $3/2$.
By examining the energies of the CDW and Fermi liquid phase in mean-field theory,
we find a first order transition from the gapped CDW phase to the Fermi liquid phase at $t/V = 0.49$, or $V/t = 2.05$.
This can be identified by the discontinuity in the order parameter.
Fig \ref{gap} shows that this is very close to the point at which the gap closes naturally in the CDW phase.


\begin{figure}
\begin{center}
\includegraphics[width=0.45\textwidth]{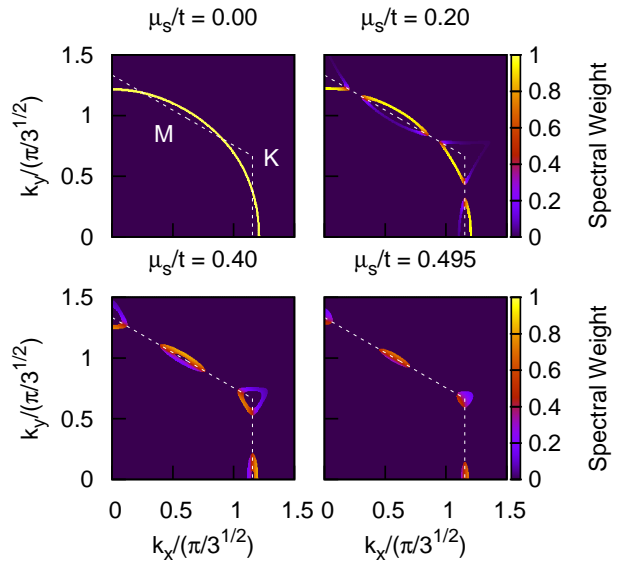}
\caption{\label{specweight}
Reconstruction of the Fermi surface when charge order is induced purely by a sublattice dependent chemical potential $\mu_s$.  
Only the first quadrant of the reduced Brillouin zone (boundary indicated by the white dotted line) is shown.
The color scale represents the spectral weight of the states along the Fermi surface.
At non-zero $\mu_s/t$, the Fermi surface is divided into pockets: particle-like surfaces along the M points, and hole-like surfaces at the K points.
The spectral weights give rise to Fermi ``arcs''.
As $\mu_s/t = 0.5$ is approached, the pockets becomes point-like, beyond which a gap is opened.
}
\end{center}
\end{figure}

\begin{figure}
\begin{center}
\includegraphics[width=0.45\textwidth]{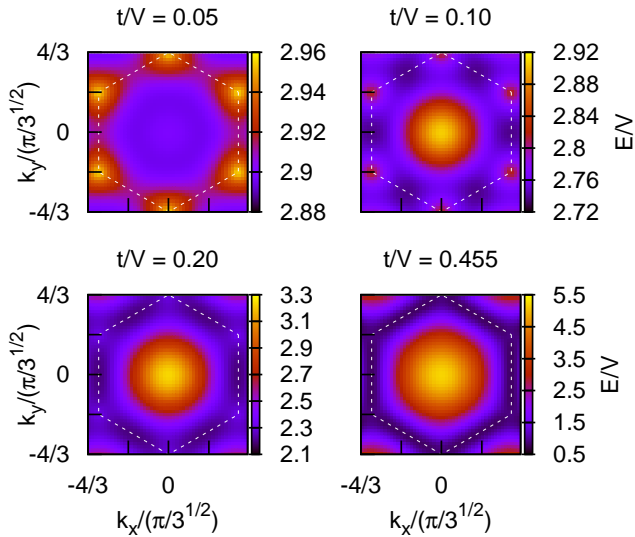}
\caption{\label{gapspec}
Series results for the spectrum of the direct gap, at $\mu_s = 0$.  
This is computed as the sum of the particle and hole excitation energies at each $k$ point.
The spectrum can be seen to change rapidly at low values of $t/V$.  
Low energy excitations can be seen to develop all along the boundary of the reduced Brillouin zone, 
as the transition at $t/V=0.49$ is approached.
This is what one expects when a gap along the boundary is created due to a small sublattice dependent chemical potential, 
suggesting that the mean-field approach is reasonable even close to the transition point.
To obtain the spectrum, we take a mean of all well-behaved differential approximants at each k-point.  
}
\end{center}
\end{figure}

Naively, one may expect the $\mu_s/t$ and $V/t$ axes of the phase diagram in Fig \ref{phasediagram} to be qualitatively similar, as the chemical potential can be thought of as a mean-field treatment of the repulsion term.
We now turn our focus to the $\mu_s/t$ axis, where $V=0$.
Along this axis, there is no first order transition.
One has first a transition from a uniform Fermi liquid at $\mu_s=0$ to a charge ordered Fermi liquid phase at low non-zero $\mu_s$.  
This can be thought of as a charge analog of the anti-ferromagnetic Fermi liquid phase \cite{sudip,palee}. 
At exactly $\mu_s/t=0.5$, an indirect gap opens up in the spectrum, although, we found no visible
singularities in the structure factor or order parameter at this point.
Thus, going along this axis represents a continuous transition from the Fermi liquid, through a charge-ordered Fermi liquid, to the gapped CDW phase.

Fig \ref{specweight} shows the reconstruction of the Fermi surface as one introduces a sublattice dependent chemical potential $\mu_s$, along with the spectral weights of the states along the surface.
At $\mu_s=0$, there exists the Fermi surface of non-interacting fermions in the full Brillouin zone.  
When a chemical potential is added, the Fermi surface undergoes a topological change, splitting into multiple disconnected pockets centered on points along the boundary of the reduced Brillouin zone.
Along the boundary, one has an alternation of particle-like surfaces centered around the M points, and hole-like surfaces around the K points.
This is a charge-ordered Fermi liquid phase, which exists at $0<\mu_s/t\leq0.5$.
At our filling, these particle and hole pockets shrink with increasing $\mu_s$, both becoming pointlike at exactly $\mu_s/t=0.5$, beyond which an indirect gap is opened and the system transitions into a gapped CDW phase.
One should note that parts of the Fermi pockets originally belonging to the non-interacting Fermi surface in the full Brillouin zone have the highest spectral weight, giving rise to Fermi ``arcs'' observable in angle resolved photoemission spectroscopy (ARPES) studies.
While this reconstruction has been thoroughly examined on the square lattice \cite{vojta,berg}, we are not aware of any studies on the triangular-lattice.

So a question that remains is: how qualitatively similar is this to the behavior along the $V/t$ axis?
Beginning in the low $t/V$ (CDW phase) limit, we see that in this region the mean-field results for energy (Fig \ref{energy}) and order parameter (Fig \ref{orderparam}) have remarkable agreement with series expansions.  
As long as the order parameter decreases monotonically and continuously with $t/V$, then the mean-field results along the $V/t$ axis is a simple mapping of those along the $\mu_s/t$ axis.
We, then, focus on the behavior near the critical point, where there is a discontinuity in the order parameter.

To confirm that a mean-field approach is reasonable close to the transition point, we investigate the direct gap spectrum obtained from series expansions.  
We examine the point at which the gap closes indirectly in the CDW phase, corresponding to a chemical potential in the mean-field treatment of $\mu_s/t = 0.5$.
The $\mu_s$ term in the Hamiltonian creates a gap between the states on the border of the reduced Brillouin zone.
In Fig \ref{gapspec}, the direct gap spectrum from series expansions shows that as the transition point is approached, the 
excitation energies become small all along the boundary of the Brillouin zone.
This is indeed consistent with what one has for non-interacting fermions when a gap is first induced by a sublattice dependent
chemical potential.

Along the $V/t$ axis, self-consistent mean-field theory shows a first order transition between the insulating CDW phase and metallic Fermi liquid phase, completely skipping any charge-ordered Fermi liquid phase.
This suggests that there does not exist any gapless long-range ordered phase along the $V/t$ axis.  
This is clearly very different than along the $\mu_s/t$ axis.
This was also found to be the case in a mean-field study of the extended Hubbard model \cite{ogata}.

\begin{figure}[t]
\begin{center}
\includegraphics[width=0.45\textwidth]{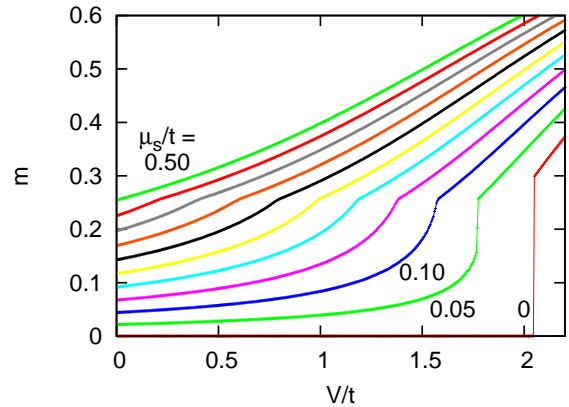}
\caption{\label{orderparam-mus}
The CDW order parameter (\ref{opdef}) as a function of $V/t$ at various values of sublattice dependent chemical potential $\mu_s/t$.  
Values of $\mu_s/t$ from $0$ to $0.50$ are shown in intervals of $0.05$.
The transition from the (charge-ordered) Fermi liquid phase at low $V/t$ to the 
gapped CDW at higher $V/t$ can be clearly seen by the discontinuity in the derivative of the order parameter, for $\mu_s/t\geq0.10$.
This suggests a second order phase transition, which occurs also at exactly the point where the gap (not shown) closes, as expected.
For $\mu_s/t\leq0.05$, there is still some discontinuity, indicating a first order transition.  
At any $\mu_s/t > 0.5$, the gap never closes, so no transition occurs.  
}
\end{center}
\end{figure}

We finally investigate the structure of the full phase diagram (Fig \ref{phasediagram}c).  
Fig \ref{orderparam-mus} shows the order parameter as a function of $V/t$ at different values of $\mu_s/t$, calculated by self-consistent mean-field theory.  
At $\mu_s/t > 0.05$, we find during the transition between the gapped CDW and the charge-ordered Fermi liquid phase,
the order parameter changes continuously, suggesting a second order transition.
The transition point corresponds to the point at which the energy gap goes continuously zero, which is expected in a second order transition.
Note that our self-consistent mean-field results, after scaling $V$ by the same factor $2/3$, show excellent agreement in the low-$t/V$ region with series expansions for energy and order parameter done at constant $\mu_s/t$ ratios (not shown).
Like in the $\mu_s=0$ case, they however disagree on the point at which the gap closes.  
We show in Fig \ref{phasediagram}c the phase boundary, identified as the closing of the gap, from mean-field and from series expansions at $\mu_s/t = 0.2$ and $\mu_s/t = 0.4$.
Series expansions show that the phase boundary is actually curved outwards, which is an aspect that self-consistent mean-field theory captures incorrectly.

Even for $\mu_s=0$, where the transition is clearly first order, it can be located 
approximately by looking for the point where the
gap appears to vanish from the CDW side.
When the transition becomes second order with non-zero $\mu_s$,  the gap closing 
defines the transition point.  
As one decreases $\mu_s$ to zero, one does not expect sudden changes in the transition point.
Thus, when the transition becomes first order, the gap closing still remains a good indicator for it.
At $\mu_s=0$, we find that the point where the energy of the Fermi liquid and CDW phase cross is very slightly different from the point where the gap closes in mean-field theory (the gap closes at roughly $V/t=1.97$, while the energy crosses at $V/t=2.05$).
This indicates that the transition does not become second order immediately with non-zero $\mu_s$, but at a small finite value.

A shortcoming of mean-field theory is its inability to accomodate short-range order.
In the Fermi liquid phase, as soon as any non-zero $V$ is introduced, short-ranged (and short-time) order will
begin to develop, but it is completely missed in mean-field theory.
It is possible that going beyond mean-field theory, short-range order or fluctuations will cause changes in
Fermi surfaces similar to those induced by a sublattice dependent chemical potential. Because our series expansion
study can not address the Fermi-liquid phase, we can not provide further insight into this issue.

In conclusion,
we have examined a $t$-$V$-$\mu_s$ spinless fermion model at $1/3$ filling on the triangular-lattice, using series expansions and self-consistent mean-field theory.  
We found that quantitative agreement of mean-field properties, such as ground state energy,
with the exact series expansions required scaling the $V$ values by an
ad hoc factor.
The Fermi liquid phase, as well as the commensurate charge-ordered Fermi liquid and gapped CDW phases are considered, and
various spectral properties are calculated.
We observe a first order transition to the insulating CDW phase at low $\mu_s$, which quickly becomes second order as $\mu_s$ is increased.
In the latter case, the gap opens gradually in the Brillouin zone, allowing for a rich reconstruction of the fermi surfaces in
the intermediate region.

Experimental evidence for such a charge density wave transition,
near $1/3$ occupancy of the triangular-lattice, has been postulated before, but not definitively observed. Fermi surface reconstruction,
similar to those observed in the cuprates may arise also in such systems. Our results are most directly applicable to single-spin
species fermion systems that may be engineered say in cold atomic gases, but may also be relevant to other strongly correlated
triangular-lattice based materials.

\begin{acknowledgements}
We would like to thank Chisa Hotta for valuable discussions.
This work is supported in part by NSF grant number  DMR-1306048.
\end{acknowledgements}


\begin{thebibliography}{2}
\bibitem{footnote1} Strictly speaking this should be called Charge Order rather than Charge Density Wave
order. However, following the common usage in the Hubbard Model literature 
(See e.g. Ref.~\onlinecite{motrunich}), we will call it Charge
Density Wave Order in this paper.
\end{thebibliography}
\end{document}